\documentclass[aps,prl,twocolumn,superscriptaddress,notitlepage]{revtex4-1}

\usepackage{epsfig}
\usepackage{amssymb} 
\usepackage{amsfonts} 
\usepackage{mathtools} 
\usepackage{dcolumn} 
\usepackage{graphicx} 
\usepackage{color}  
\usepackage{bm}  
\usepackage{comment} 
\usepackage{units}
\usepackage{sidecap} 
\usepackage{textcomp} 
\usepackage{grffile} 
\usepackage{natbib}
\usepackage{soul}
\usepackage{gensymb}

\usepackage{array} 

\newcommand{\bpm}{\begin{pmatrix}}
\newcommand{\epm}{\end{pmatrix}}
\newcommand{\be}{\begin{eqnarray}}
\newcommand{\ee}{\end{eqnarray}}
\newcommand{\ba}{\begin{array}}
\newcommand{\ea}{\end{array}}

\begin{document}
\title{Thermodynamic signatures for the existence of Dirac electrons in ZrTe$_5$}

\author{Nityan L.~Nair}
\affiliation{Department of Physics, University of California, Berkeley, California 94720, USA}
\affiliation{Materials Sciences Division, Lawrence Berkeley National Laboratory, Berkeley, California 94720, USA}

\author{Philipp T.~Dumitrescu}
\affiliation{Department of Physics, University of Texas at Austin, Austin, TX 78712, USA}

\author{Sanyum Channa}
\affiliation{Department of Physics, University of California, Berkeley, California 94720, USA}

\author{Sin\'{e}ad M. Griffin}
\affiliation{Department of Physics, University of California, Berkeley, California 94720, USA}
\affiliation{Molecular Foundry, Lawrence Berkeley National Laboratory, Berkeley, CA 94720, USA}

\author{Jeffrey B. Neaton}
\affiliation{Department of Physics, University of California, Berkeley, California 94720, USA}
\affiliation{Molecular Foundry, Lawrence Berkeley National Laboratory, Berkeley, CA 94720, USA}
\affiliation{Kavli Energy NanoScience Institute at Berkeley, Berkeley, CA 94720, USA}
\affiliation{Materials Sciences Division, Lawrence Berkeley National Laboratory, Berkeley, California 94720, USA}

\author{Andrew C.~Potter}
\affiliation{Department of Physics, University of Texas at Austin, Austin, TX 78712, USA}

\author{James G.~Analytis}
\affiliation{Department of Physics, University of California, Berkeley, California 94720, USA}
\affiliation{Materials Sciences Division, Lawrence Berkeley National Laboratory, Berkeley, California 94720, USA}

\date{\today}

\begin{abstract}
We combine transport, magnetization, and torque magnetometry measurements to investigate the electronic structure of ZrTe$_5$ and its evolution with temperature. At fields beyond the quantum limit, we observe a magnetization reversal from paramagnetic to diamagnetic response, which is characteristic of a Dirac semi-metal. We also observe a strong non-linearity in the magnetization that suggests the presence of additional low-lying carriers from other low-energy bands. Finally, we observe a striking sensitivity of the magnetic reversal to temperature that is not readily explained by simple band-structure models, but may be connected to a temperature dependent Lifshitz transition proposed to exist in this material.
\end{abstract}

\maketitle

Thermodynamic signatures of the topological nature of a material have become increasingly important as numerous new topological insulator, Weyl, and Dirac materials are predicted \cite{Chadov2010, Ashwin, Cava}. In this study we focus on ZrTe$_5$, a material whose topological nature is hotly debated; it has been predicted and verified as a Dirac semimetal \cite{chen_magnetoinfrared_2015,Li2016,Zheng2016, yuan_observation_2016,liu_zeeman_2016}, a topological insulator \cite{Chen2017,li_experimental_2016,wu_evidence_2016,manzoni_evidence_2016} and a trivial semiconductor~\cite{ZrTe5_trivial, moreschini_nature_2016}. In addition to its potential topological nature, there is also an unusual anomaly in the temperature dependence of the resistivity that has been conjectured to originate from a Lifshitz transition \cite{ZrTe5_anomaly,Chi2017}. This anomaly is strongly sample dependent, ranging in its position from $\sim$ 10K to 150K. Recently, it has been suggested that the topological nature of the band structure and associated transport properties of ZrTe$_5$ depend strongly on the growth technique~\cite{fan_transition_2017}. Nevertheless, the complicated history of this material highlights the need for robust low-energy signatures of Dirac-like band structures.

We study the magnetic behavior of ZrTe$_5$ as it approaches and surpasses its quantum limit - the magnetic field at which all electrons are collapsed into the lowest, $\nu = 0$ Landau level. In the most general case, all materials show some small degree of orbital diamagnetism arising from the local orbital moment of the ions. Trivial metals show an additional Landau diamagnetism arising from the orbital motion of their itinerant electrons. In a previous study of NbAs \cite{Moll2016}, we showed that topological metals exhibit a low-field paramagnetic response originating from their unique Landau quantization, which for a Dirac fermion in a magnetic field $B$ along $z$ is given by
\begin{equation}
\varepsilon_{\nu,k} = \hbar v_F \sqrt{2B(\nu+\gamma) + k_z^2},
\label{eq:LL}
\end{equation} 
where $\nu$ is the Landau index, $v_F$ the Fermi velocity and $\gamma$ is the quantum correction term which is $1/2$ for trivial metals, but in Dirac systems $\gamma=0$ due to the non-trivial Berry's phase. The Berry's phase associated with this quantization is often used as evidence for the existence of non-trivial topology, which can in principle be extracted from a plot of the Landau indices versus inverse magnetic field \cite{zhang_experimental_2005}. However, the influence of Zeeman splitting, the complicating effects of conductivity contributions from other bands and the presence of a Dirac mass can make this extraction unreliable, particularly in three dimensions \cite{roth_semiclassical_1966,wright_quantum_2013}. Instead, a robust consequence of Eq. \ref{eq:LL} is a sign-change of the magnetization beyond the quantum limit, which in the absence of any other magnetic transitions, is a characteristic signature of the presence of an electron-like Weyl or Dirac Fermi surface. Because $\gamma = 0$ in Eq. \ref{eq:LL}, the $\nu=0$ Landau level is pinned at zero energy. As the magnetic field is increased, the growing degeneracy of this Landau level pulls down the total energy of the system, leading to an overall paramagnetic response given by $M = -dE/dB$. When the quantum limit is exceeded, the chemical potential asymptotically approaches the $\nu=0$ Landau level and the paramagnetic response diminishes, crossing over to the diamagnetic response expected from fully occupied bands~\cite{Moll2016}. This shift leads to a kink at the quantum limit which, when combined with the sign change, provides strong evidence for the Dirac nature of the system.

\begin{figure*}
 \begin{center}
   \includegraphics[width=1\textwidth]{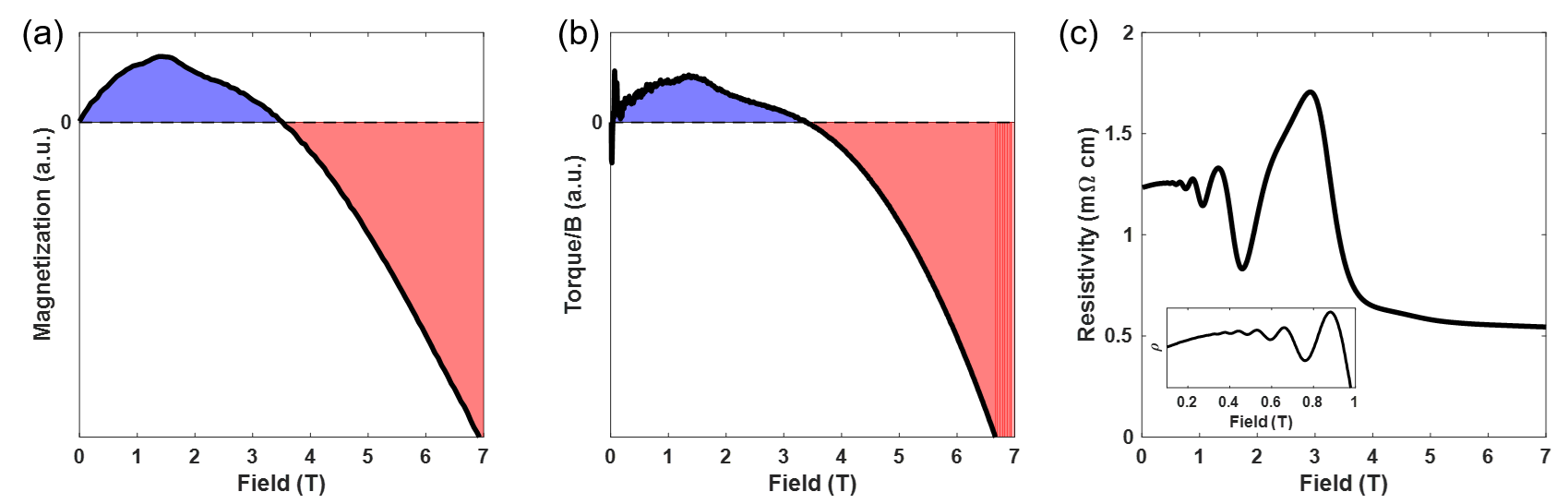}
\end{center}
  \caption{\small{({a}) The magnetization of ZrTe$_5$ with magnetic field applied along the crystallographic b-axis shows a clear paramagnetic response at low field and a transition to a diamagnetic response at high field.  ({b}) The magnetic torque measured on the same sample as ({a}) is in close agreement with the magnetization and exhibits the same transition from para- to dia-magnetism. ({c}) The magnetoresistance of ZrTe$_5$ shows pronounced SdH oscillations and the onset of the quantum limit in the vicinity of the sign change observed in ({a}) and ({b}). Inset: The low-field oscillations show no evidence of beating, implying that only one spin-split frequency is being observed. All measurements were taken at 1.8K.}}
\label{fig:1}
\end{figure*}

\begin{figure*}
 \begin{center}
   \includegraphics[width=1\textwidth]{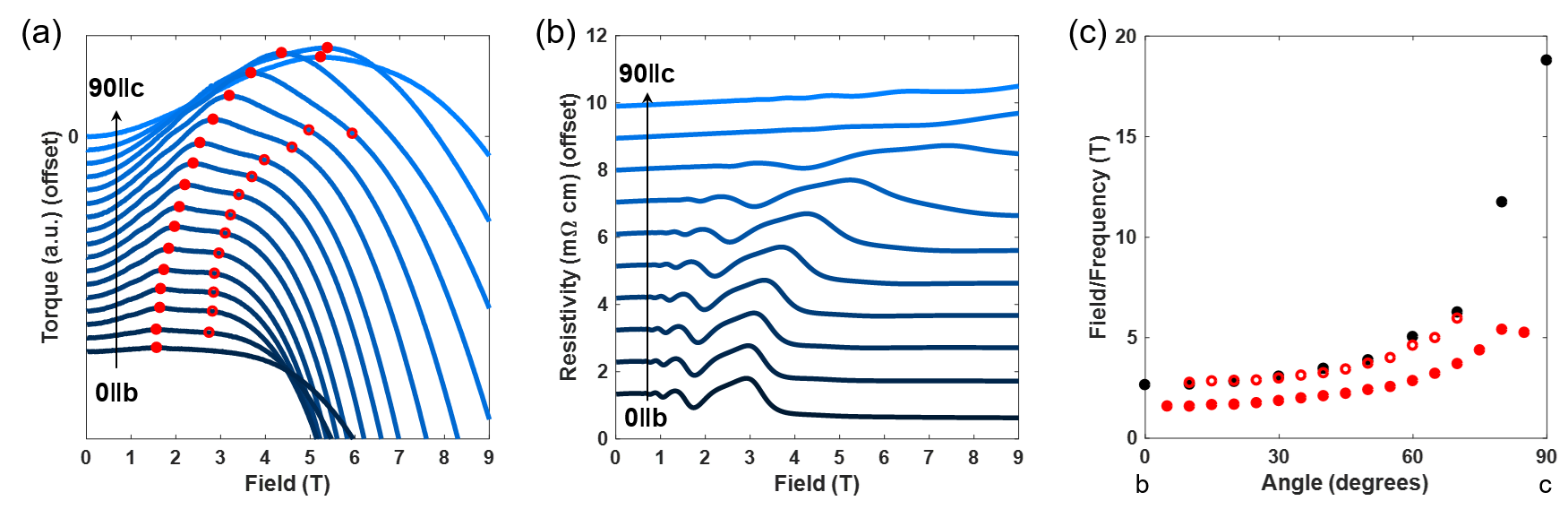}
\end{center}
  \caption{\small{ ({a}) Torque measured at different magnetic field orientations in the b-c plane. The paramagnetic response and cross-over field grows with field angle. Two kinks can be extracted from the data. The more prominent low-field kink (filled circles) can be tracked for all angles. The less prominent high-field kink (empty circles) is only observable for intermediate angles. ({b}) The magnetoresistance measured at different field orientations in the b-c plane shows that the quantum limit grows monotonically with increasing field angle. ({c}) The kinks in the magnetic torque from ({a}) (red circles) compared to the SdH oscillation frequency extracted from the magnetoresistance data in ({b}) (black circles). The quantum limit, which occurs at the SdH frequency, tracks well with the high-field kink over the observable range. All measurements were taken at 1.8K}}
\label{fig:2}
\end{figure*}

ZrTe$_5$ single crystals were grown by the vapor transport technique using iodine as the transport agent. Precursor powder was prepared by sealing a stoichiometric mixture of Zr and Te in a quartz ampule under vacuum which was heated to 500$^\circ$C and held for 7 days. The resulting powder was mixed with 5mg/cm$^3$ of iodine and sealed in a quartz ampule under vacuum before being loaded into a two-zone furnace. The source and sink ends of the ampule were held at 520$^\circ$C and 480$^\circ$C, respectively for 21 days. Needle-like crystals up to 7mm long were obtained from the cold end of the ampule. Crystal structure and orientation was confirmed using x-ray diffraction.

In Fig. \ref{fig:1}A-C we illustrate the main result of this work. The magnetization (Fig. \ref{fig:1}A) is observed to be paramagnetic at low field crossing over to diamagnetic at high field. The magnetic torque, related to the magnetization by $\vec{\tau} = \mu_0 \vec{M}\times \vec{H}$, shows a very similar behavior. Both exhibit two kinks: a dominant kink at $\sim1.5T$ and a smaller kink at $\sim2.5T$, after which the sign change occurs. Finally, Fig. \ref{fig:1}C shows the magnetoresistance of ZrTe$_5$, where pronounced oscillations are observable from the Shubnikov-de Haas (SdH) effect. Note that the quantum oscillations are spin-split, so the Landau level position should be identified as the trough between spin-split SdH peaks. From this we can identify that the $\nu = 1$ Landau level crosses the Fermi energy at $2.6T$, in agreement with the frequency extracted from the low-field SdH oscillations and very close to the smaller kink at $\sim 2.5T$. We conclude that the smaller kink corresponds to the Dirac pocket entering the quantum limit and the associated loss of paramagnetism. We believe that the detailed size and shape of the magnetization, including the larger kink, is determined by the presence of additional bands, as we will discuss below.

In order to distinguish the low-field behavior from other sources of paramagnetism we confirm that the sign change in the magnetization tracks with the quantum limit of the Fermi surface by comparing the angle dependence of the torque signal and the SdH oscillations (Fig.\ref{fig:2}). As the angle of the field with respect to the principal axes of the crystal is changed, the cross-sectional area tracked by the frequency of the SdH oscillations shifts accordingly, pushing the quantum limit out to higher field. Fig. \ref{fig:2}B shows SdH oscillations in the magnetoresistance of ZrTe$_5$ at different angles. The frequency and the expected position of the $\nu=1$ Landau level can be extracted from the oscillation and is plotted in Fig. \ref{fig:2}C (black circles).  The anomaly observed in the torque (empty circles) tracks perfectly with the expected position of the $\nu = 1$ Landau level up until angles where the kink is so broad that it can no-longer be tracked. However, we can track the $\nu = 2$ feature (filled circles in Fig. \ref{fig:2}A) in the torque to much higher fields, and it can be seen to follow the angle dependence of the SdH very closely up to high angles. The correlation between the kink in magnetic torque and loss of paramagnetism with the quantum limit provides unambiguous evidence for the presence of an electron-like Dirac Fermi surface which is responsible for the observed SdH oscillations.

\begin{figure}
 \begin{center}
   \includegraphics[]{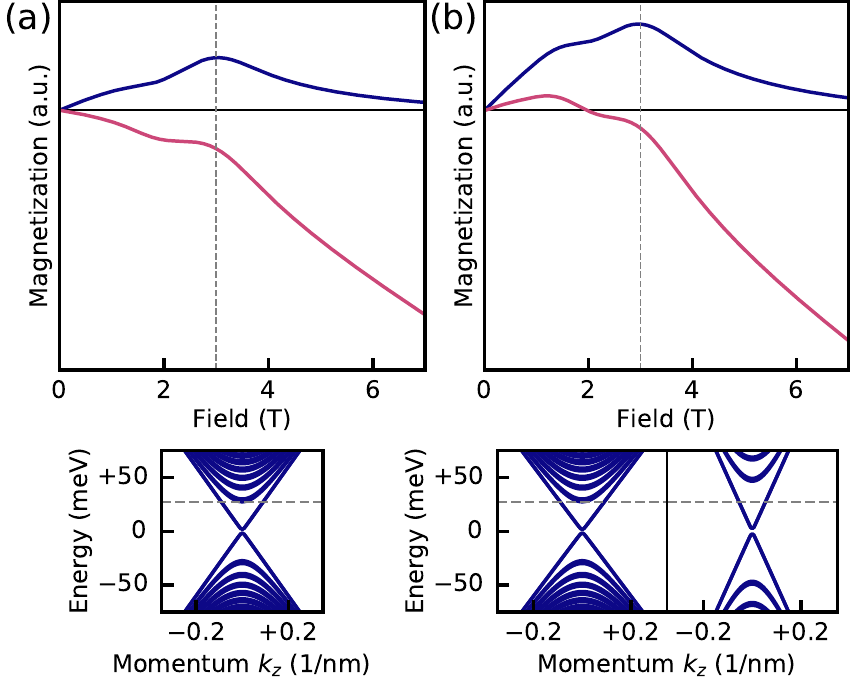}

\end{center}

\caption{\small{(a) Upper: Simulated magnetization of a single Dirac band at constant density, with (red) and without (blue) a linear diamagnetic background. The vertical dashed line at $3~\mathrm{T}$ indicates where the Dirac band enters its quantum limit. Lower: Dirac band-structure at magnetic field $3~\mathrm{T}$ with chemical potential $\mu$ indicated by the grey dashed line.
  (b) Simulated magnetization for two Dirac bands with (red) and without (blue) a linear diamagnetic background. The response from the high velocity Dirac pocket enhances the paramagnetic response at low field; with background diamagnetism, the $\nu=2$ peak may become dominant.}}
\label{fig:4}
\end{figure}

There is one peculiar feature of our data which requires additional explanation. Although the $\nu=1$ kink observed in the magnetic torque tracks well with the quantum limit, the dominant change in curvature of the torque signal appears at magnetic fields lower than the quantum limit (see the filled circles in Fig. \ref{fig:2}A). Figure \ref{fig:4}a shows a simulation of the magnetization from a single Dirac conduction band with a linear diamagnetic background (see Supplementary Information for details). While this broadly reproduces the oscillatory part of the data, it does not capture the overall curvature of the magnetization and the position of the dominant kink. In particular, to reduce the magnitude of the feature arising from the $\nu=1$ Landau Level below that of the $\nu=2$ with linear diamagnetism only, one has to suppresses the weak field paramagnetism entirely. This implies that a non-linear diamagnetic background must be present.

While there are a number of possible mechanisms for an additional non-linear response, an appealing possibility is the presence of second Dirac pocket. This pocket -- which may be itself slightly gapped -- would enter its quantum limit at weak fields and enhance the low field paramagnetic signal. As shown in Fig.~\ref{fig:4}b, a simulation of this system better captures the key features of the data. We note that if another Dirac pocket exists, we observe no direct signature in the quantum oscillatory signal. Nevertheless, the existence of such a pocket is consistent with DFT calculations of the band structure (see Supplementary Information), and may have already been observed in ARPES experiments \cite{moreschini_nature_2016, zhang_electronic_2017}. 

Our DFT calculations find several features close to the Fermi level as shown in the band structure calculation in the Supplementary Information. Without including spin-orbit coupling we find two Dirac crossings, close to the $\Gamma$ and S points. With the inclusion of spin-orbit interaction, the Dirac crossings become gapped and a new feature with a Dirac-like dispersion appears at $\Gamma$, in addition to several other bands close to the Fermi level in the Z to T direction. Hall effect data shows electron-like carriers at low temperature \cite{Chi2017}, implying some of these additional bands must be populated and Dirac-like or massive Dirac-like features should be present. Note that these features are extremely sensitive to cell volume and strain \cite{fan_transition_2017}, which may explain the conflicting experimental reports on the electronic properties and topological signatures in ZrTe$_{5}$ \cite{chen_magnetoinfrared_2015,Li2016,Zheng2016, yuan_observation_2016,liu_zeeman_2016, Chen2017,li_experimental_2016,wu_evidence_2016,manzoni_evidence_2016, ZrTe5_trivial, moreschini_nature_2016}.

\begin{figure}
 \begin{center}
   \includegraphics[width=0.4\textwidth]{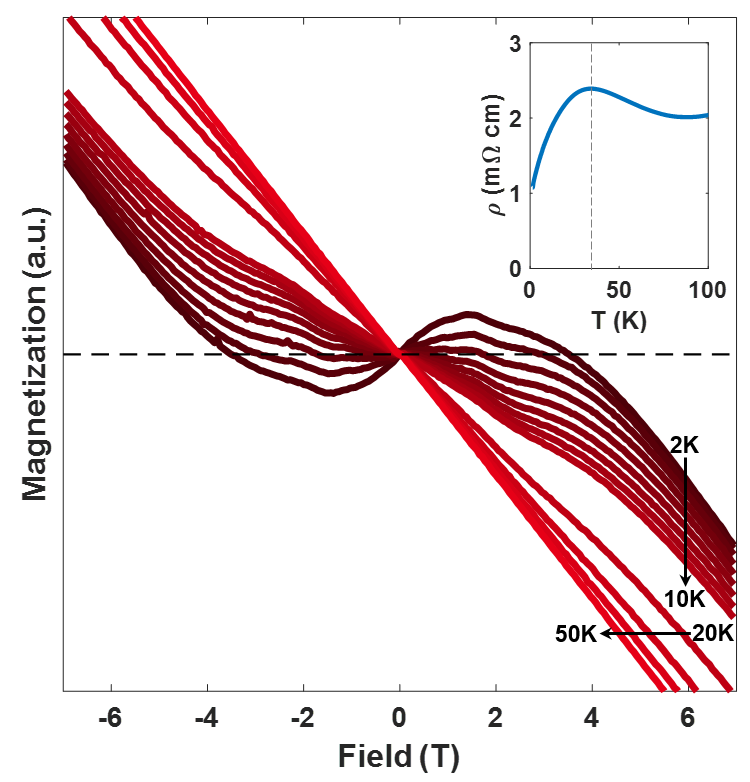}
\end{center}
  \caption{\small{Temperature dependence of the magnetization with magnetic field oriented along the crystallographic b-axis. The low-field paramagnetic response is rapidly suppressed with increasing temperature, becoming completely diamagnetic by 5K. Above 30K, the magnetization approaches the constant, temperature-independent diamagnetic response typically found in ordinary metals. Inset: The resistivity shows a peak around 30K, which has been attributed to a Lifshitz transition in ZrTe$_5$ and matches the temperature scale at which the paramagnetism disappears.}}
\label{fig:3}
\end{figure}

We now turn to the temperature dependence of the magnetic signal, shown in Fig. \ref{fig:3}. Strikingly, the low-field paramagnetic response is rapidly suppressed by increasing temperature and completely disappears by 5K. This suppression strongly suggests that the balance of Dirac and non-Dirac contributions to the total magnetization is highly dependent on thermal processes. By 10K the magnetization is dominated by the diamagnetic response, but its non-linearity suggests there is still a competing contribution from the Dirac pocket. At temperatures above 30K (which coincides with the peak observed in transport), the magnetization approaches the temperature-independent diamagnetic response typical of ordinary materials. In general, such a sign reversal would not be expected in a thermodynamic quantity over such a short temperature range without a phase transition. This surprising result indicates that the Dirac-like signatures of ZrTe$_5$ are very sensitive to low energy processes.

We note that the disappearance of the paramagnetic response coincides with a peak in the resistivity occurring at 35K. This peak, which has been observed from ~10K to 150K in samples grown by other groups, has been attributed to a Lifshitz transition \cite{Chi2017}. The coincidence of the magnetization's temperature dependence with the resistivity hump strongly suggests that the two originate from the same mechanism: either the addition of trivial carriers to suppress the Dirac contribution to the magnetization or the loss of Dirac carriers. Although we do not have direct evidence to support either, the fact that the resistivity shows a peak instead of a trough weighs in favor of the latter.

To summarize, by combining torque, magnetization and magneto-transport data we have provided strong evidence that ZrTe$_5$ is a Dirac semimetal. Interestingly, the thermodynamic signature of its Dirac nature vanishes quickly with increasing temperature and is apparently undetectable at temperatures above the resistivity hump. This anomaly has been associated with a Lifshitz transition of a Dirac pocket as discussed by other authors, and while it is possible that other, more complex mechanisms could explain our data, this picture seems consistent with our data. This study thus provides a direct thermodynamic signature of the presence of Dirac fermions in ZrTe$_5$ and potentially the first magnetic signature of a Dirac Lifshitz transition.

\section{Acknowledgments}
This work was supported by the Gordon and Betty Moore Foundation's EPiQS Initiative through Grant No. GBMF4374 and the Office of Naval Research under the Electrical Sensors and Network Research Division Award No. N00014-15-2674. N. L. N. was supported by NSF GRFP under Grant No. DGE 1106400. S. M. G. and J. B. N. were supported by the Director, Office of Science, Office of Basic Energy Sciences, Materials Sciences and Engineering Division, of the U.S. Department of Energy under Contract No. DE-AC02-05-CH11231. Computational resources provided in part by the Molecular Foundry was supported by the Office of Science, Office of Basic Energy Sciences, of the U.S. Department of Energy, also under Contract No.
DE-AC02-05-CH11231. The authors would like to thank Camelia Stan for her help in performing crystal diffraction and orientation measurements on beamline 12.3.2 at the Advanced Light Source. The Advanced Light Source is a DOE Office of Science User Facility supported under contract no. DE-AC02-05CH11231.

%

\end{document}